\documentclass[pra,twocolumn,showpacs,superscriptaddress,floatfix]{revtex4-1}
\usepackage{amsmath}
\usepackage{amsfonts}
\usepackage{amssymb}
\usepackage{mathrsfs}
\usepackage{latexsym}
\usepackage{epstopdf}
\usepackage{dsfont}
\usepackage{graphicx}
\usepackage[unicode=true, pdfusetitle,
bookmarks=true,bookmarksnumbered=false,bookmarksopen=false,
breaklinks=false,pdfborder={0 0 0},backref=false,colorlinks=true]
{hyperref}
\usepackage{epstopdf}

\begin{document}

\title{Witnessing spin-orbit thermal entanglement in rare-earth ions} 

\author{O. S. \surname{Duarte}}
\email{oduartem@if.uff.br}
\affiliation{Instituto de F\'{\i}sica, Universidade Federal Fluminense, Av. Gal. Milton Tavares de Souza s/n, Gragoat\'a, 24210-346, Niter\'oi, RJ, Brazil.}

\author{C. S. \surname{Castro}}
\email{ccastro@if.uff.br}
\affiliation{Instituto de F\'{\i}sica, Universidade Federal Fluminense, Av. Gal. Milton Tavares de Souza s/n, Gragoat\'a, 24210-346, Niter\'oi, RJ, Brazil.}

\author{D. O. Soares-Pinto}
\affiliation{Instituto de F\'{i}sica de S\~{a}o Carlos, Universidade de S\~{a}o Paulo, P.O. Box 369, S\~{a}o Carlos, 13560-970 SP, Brazil}

\author{M. S. \surname{Reis}}
\affiliation{Instituto de F\'{\i}sica, Universidade Federal Fluminense, Av. Gal. Milton Tavares de Souza s/n, Gragoat\'a, 24210-346, Niter\'oi, RJ, Brazil.}

\date{\today}

\begin{abstract}
We explore spin-orbit thermal entanglement in rare-earth ions, based on a witness obtained from mean energies. The entanglement temperature $T_{E}$, below which entanglement emerges, is found to be thousands of kelvin above room temperature for all light rare earths. This demonstrate the robustness to environmental fluctuations of entanglement between internal degrees of freedom of a single ion.
\end{abstract}

\maketitle

The phenomenon of entanglement has received much attention from the scientific community owing to its importance in the foundations of Quantum Mechanics and the possibility of its use as a resource in quantum information processesing, for example, in dense coding \cite{1992_PRL_69_2881} and quantum teleportation \cite{1993_PRL_70_1895}. In spite of the advances in this field~\cite{2009_RMP_81_865}, some important open questions still remain, such as how to determine in an unique and universal way whether a general mixed state is entangled or not in relation to a particular partition. Thus, a large portion of the research in quantum information theory is concerned with developing tools capable of detecting or, at least, estimating entanglement \cite{2007_NJP_9_46, 2007_PRL_98_110502, 2005_PRA_72_032309}. Among these tools, a very useful one, which can estimate the degree of entanglement in states of arbitrary purity and dimensions, even in many-body systems \cite{2008_RMP_80_517}, is a class of observables called \textit{entanglement witnesses} \cite{1996_PLA_223_1, 2000_PLA_271_319}. Such observables present positive expectation values for all separable states, and negative expectation values when entanglement is present~\cite{2009_PhysRep_474_1}. 

Another important question is whether entanglement can survive at relatively high temperatures, although the process of decoherence increases with the size and the temperature of the system~\cite{2001_PRL_87_017901,2011_PRB_84_214431,2009_PRL_102_100503,2008_PRL_100_080502,2008_EPL_81_40006}. Unlike previous studies~\cite{nature_425_2003_48,2006_PRB_73_134404,2007_PRB_75_054422,2008_PRB_77_104402,2009_PRB_79_054408,2009_EPL_87_40008,2012_EPL_100_50001}, which look for entanglement between angular momenta of the same nature (namely, spins) in different ions, the present study focuses attention on entanglement between internal degrees of freedom of a single ion, spin and orbital angular momenta~\cite{2012_JPCM_24_313201}, at finite temperature. This approach shows that rare-earth ions are good candidates to retain entanglement up to high temperatures, due to the huge energy gap, provided by spin-orbit coupling, between the ground state and the first excited state. In order to determine the spin-orbit thermal entanglement, we use a technique analogous to that described by T\'{o}th~\cite{2005_PRA_71_010301}, in which a witness that is directly related to the mean energies~\cite{2012_PRA_86_032330} signals the presence of absence of entanglement. When the system is in thermal equilibrium, the internal energy can be used as an entanglement witness, and a temperature bound for separable states can be found.


Rare-earth ions present an incomplete $4f$ shell (except for La and Lu). When forming compounds they lose $6s$ and $4f$ electrons, and the remaining unpaired $4f$ electrons are responsible for their magnetic properties. Since $4f$ shells are spatially inside $5s$ and $5p$ shells, a shield effect is present and the $4f$ electrons are isolated from their surroundings~\cite{livro_jensen}. The magnetic behavior in terms of localized magnetic moments is related to an effective spin operator $\textbf{S}$, with associated eigenvalue $s$, and an effective orbital angular momentum operator $\textbf{L}$, with associated eigenvalue $l$, given by the Hund rules. These two quantities compose the total angular momentum $\textbf{J} = \textbf{L} + \textbf{S}$, with eigenvalues $|s-l|\leq j \leq s+l$.

Thus the Hamiltonian describing the interaction between these two quantities is
\begin{equation}
H = \zeta\,\textbf{S}\cdot\textbf{L}
\label{H}
\end{equation}
where $\zeta$ is the spin-orbit coupling parameter. Depending on the value of the coupling parameter, two different scenarios arise. If the $4f$ shell is less than half filled, as it is in the light rare-earth ions, $\zeta$ is positive and therefore the coupling between the vectors $\textbf{S}$ and $\textbf{L}$ is anti-parallel. Consequently, the ground state is characterized by the total angular momentum $j_{0}=\left|s-l\right|$. If the $4f$ shell is more than half filled, as it is in the heavy rare-earth ions, $\zeta$ is negative and therefore the coupling is parallel. As before, the ground state is then given by $j_{0}=s+l$.  The values of $s$, $l$ and $j_{0}$ for the rare-earth ions are given in Table~\ref{values1}.

\begin{table*}
\begin{centering}
\begin{tabular}{c|c|c|c|c|c|c|c|c|c|c|c|c|c|c}
\hline 
    & Ce & Pr & Nd & Pm & Sm & Eu &Gd & Tb & Dy & Ho & Er & Tm & Yb\tabularnewline
\hline
\hline
$s$ & 1/2 & 1 & 3/2 & 2 & 5/2 & 3 & 7/2 & 3 & 5/2 & 2 & 3/2 & 1 & 1/2\tabularnewline
\hline
$l$ & 3 & 5 & 6 & 6 & 5 & 3 & 0 & 3 & 5 & 6 & 6 & 5 & 3\tabularnewline
\hline
$j_{0}$ & 5/2 & 4 & 9/2 & 4 & 5/2 & 0 & 7/2 & 6 & 15/2 & 8 & 15/2 & 6 & 7/2\tabularnewline
\hline
$\Delta\mathcal{E}$(K) & 3150 & 3100 & 2750 & 2300 & 1450 & 500 & 43200 & 2900 & 4750 & 7500 & 9350 & 11950 & 14800 \tabularnewline
\hline
$\zeta$(K) & 900 & 620 & 500 & 460 & 414 & 500 &---& $-483$ & $-633$ & $-937$ & $-1247$ & $-1991$ & $-4229$ \tabularnewline
\hline
$T_{E}$(K) & 1758 & 1851 & 1904 & 2008 & 1975 & 3295 &--- &---&---&---&---&---&---\tabularnewline
\hline
\end{tabular}
\par\end{centering}
\caption{\label{values1} Values of spin $s$, orbital angular momentum $l$, ground state total angular momentum $j_{0}$, energy gap $\Delta\mathcal{E}$, coupling parameter $\zeta$ and entanglement temperature $T_{E}$ for rare-earth ions.}
\end{table*}

For the sake of completeness, let us rewrite the Hamiltonian in Eq.(\ref{H}) as
\begin{equation}
H = \frac{\zeta}{2}\left( J^{2} - S^{2} - L^{2} \right),
\end{equation}
whose energies are 
\begin{equation}
\mathcal{E}_{j} = \frac{\zeta}{2} \left[ j(j+1) - s(s+1) - l(l+1) \right].
\label{energy}
\end{equation}
Thus, from Eq.(\ref{energy}), we can see that the energy of each multiplet $j$ is proportional to the coupling parameter $\zeta$. It is worth noting that the coupling parameter reaches thousands of Kelvins, as can be seen in Table \ref{values1}, meaning a huge energy gap between the ground state and first excited state.


The motivation for studying light rare earths was the anti-parallel alignment between angular momenta. If both angular momenta of the pair present eigenvalue $\frac{1}{2}$, which means a $2\times2$ Hilbert space with anti-parallel coupling, the explanation for entanglement is quite obvious, since there are only the singlet and triplet eigenstates. In this case, the singlet corresponds to the ground state and the entanglement between the variables is easily determined, since it is a pure state. On the other hand, as the triplet is formed by a mixture of separable and entangled states, no entangled can be found. If the coupling were parallel the situation would become the opposite, with the triplet being the ground state and, consequently, no entanglement is present \cite{2002_PLA_301_1}.

To study entanglement between spin and orbital angular momenta for a rare-earth ions,  it is not enough to check directly the eigenstates of the Hamiltonian of Eq.(\ref{H}) for a given value of $l$ and $s$. Since the angular momenta are of different nature and size, the Hilbert space of the system has a dimension $d_{s}\otimes d_{l}$ and there is no unique criterion for determining entanglement in mixed states in such circumstances. In fact, the only pure ground state is the one from the light rare-earth ion Europium, all the other ions having degenerate ground states. Thus it would be useful to have an observable capable of revealing entanglement without reference to the eigenstates of Eq.(\ref{H}).




Following the general idea given in Ref.\cite{2005_PRA_71_010301}, to construct an observable capable of witnessing entanglement between spin and orbital angular momenta, we consider the ensemble average of Eq.(\ref{H}) related to a general product state $\rho=\rho_{S}\otimes\rho_{L}$, where $\rho_{S}$ and $\rho_{L}$ are density operators in the Hilbert space of spin and orbital angular momentum, respectively. Thus 
\begin{eqnarray}
\left\langle H\right\rangle _{\rho} &=&\zeta\left\langle \textbf{S}\right\rangle _{\rho_{S}}\cdot\left\langle \textbf{L}\right\rangle _{\rho_{L}} \nonumber \\
&=&\zeta\left|\left\langle \textbf{S}\right\rangle _{\rho_{S}}\right|\cdot\left|\left\langle \textbf{L}\right\rangle_{\rho_{L}}\right|\cos\theta,\label{Av(H)_sep}
\end{eqnarray}
where $\left| \textbf{A}\right|$ represents the magnitude of a vector $\textbf{A}$ and $\theta$ is the angle between the vectors $\left\langle \textbf{S}\right\rangle$ and $\left\langle \textbf{L}\right\rangle$. 

The minimization of $\left\langle H\right\rangle _{\rho}$ depends on the signal of $\zeta$. For $\zeta>0$ (light rare earths) the infimum is attained when $\theta=\pi$, and for $\zeta<0$ (heavy rare earths), the angle $\theta=0$ minimizes $\left\langle H\right\rangle _{\rho}$. In addition, for a state of angular momentum $j$, the inequality $\left|\left\langle \textbf{J}\right\rangle\right| \leq j$ holds. Thus the infimum of Eq.(\ref{Av(H)_sep}) is $\mp\zeta sl$, where the upper (lower) sign is for positive (negative) $\zeta$. Then, for a general product state, the lower bound
\begin{equation}\label{H_bound}
\left\langle H\right\rangle _{\rho}\geq\mp\zeta sl
\end{equation}
is satisfied. Finaly, the internal energy can be used as an entanglement witness and the following relation must hold for all separable thermal states:
\begin{equation}
\mathcal{W}_{H}\equiv\left\langle H\right\rangle \pm\zeta sl\geq0,\label{Wh}
\end{equation}
where $\left\langle H\right\rangle =\mbox{tr}\{e^{-\beta H}H\}/Z$, with $\beta=(k_{B}T)^{-1}$ and $Z=\mbox{tr}\{e^{-\beta H}\}$.

\begin{figure}[h!]
\begin{center}
\includegraphics[scale=0.6]{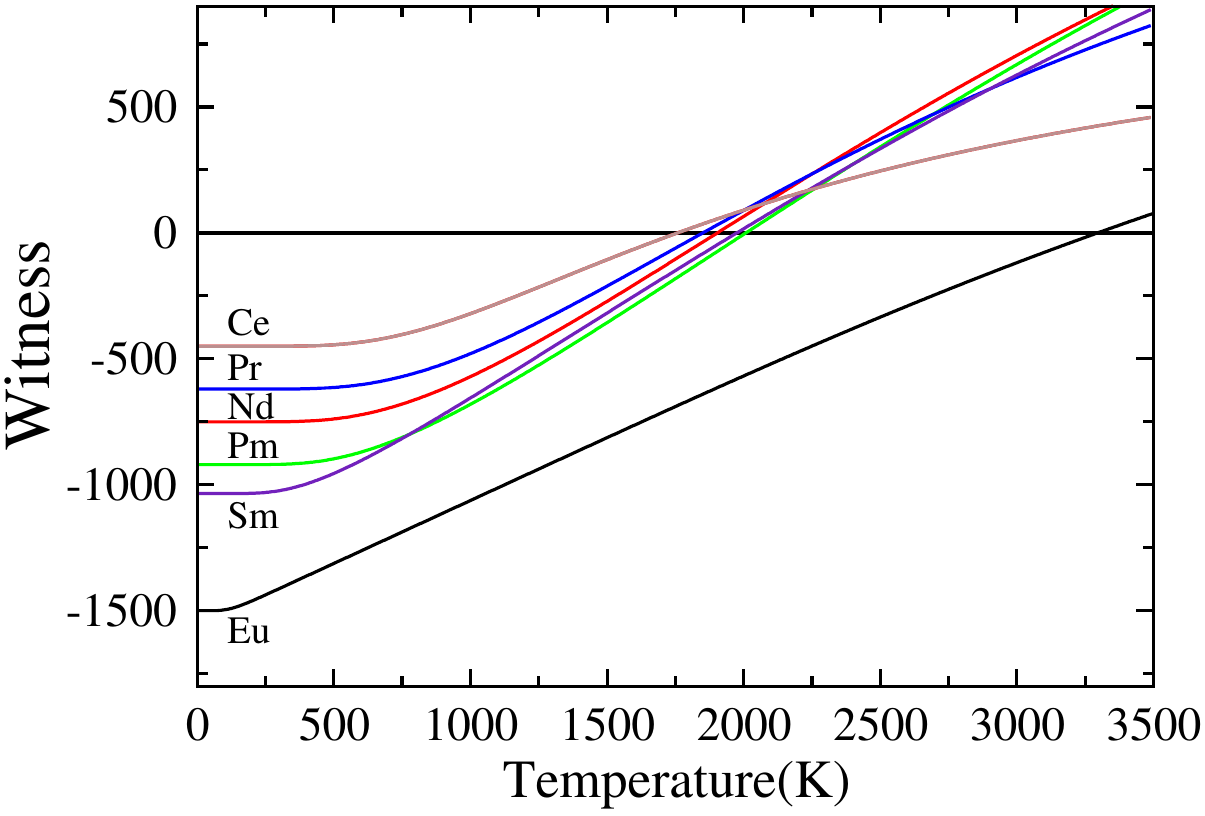}
\end{center}
\caption{(color online).\label{fig:AllRE2} Entanglement witness for light rare-earth ions as a function of temperature. As can be seen, all light rare-earth ions present entanglement up to high temperatures.}
\end{figure}

The equality in Eq.(\ref{Wh}) allows the definition of a temperature $T_{E}$, below which the state will be certainly entangled with respect to the $S-L$ bi-partition. Figure \ref{fig:AllRE2} presents the entanglement witness as a function of temperature for light rare earths. The coupling parameter $\zeta$ was obtained from the energy difference $\Delta\mathcal{E}=\zeta (j+1)$ for light rare earths and $\Delta\mathcal{E}=-\zeta j$ for heavy rare earths. Table \ref{values1} shows the highest values of temperature below which there exists spin-orbit entanglement, the values of $\zeta$ for each rare-earth ion and the energy gap $\Delta\mathcal{E}$. The light rare earths present spin-orbit entanglement for temperatures up to 3295 K, but no entanglement is observed for heavy rare earths.

Figure \ref{fig:AllRE2} shows the entanglement for all the light rare earths. The purity of Europium's ground state makes entanglement more robust to rising temperature. As we can see from Table \ref{values1}, Europium is the only rare earth that reaches an entanglement temperature much higher (more than six times higher) than its excitation gap $\Delta\mathcal{E}$. In the other cases, even with degenerate ground states, entanglement is present and it survives up to temperatures much higher than room temperature, but, except for Samarium, the temperatures are lower than the energy gap.

The results for heavy rare earths are completely different. First of all, the witness does not predict entanglement in any of the heavy rare earths. Because of the size of the angular momenta involved, the explanation is not direct as in the case of pairs with eigenvalue $\frac{1}{2}$. However, from the energy spectrum it is evident that there exists an inversion of eigenvectors when compare with the light rare earths. That is, for given values of $s$ and $l$, we have a light and a heavy rare earth with the same energy eigenvectors, but the eigenvalues are organized in ascending order from the lowest to the highest, with the $j$-label going from $|l-s|$ to $l+s$ for light rare earths and from $l+s$ to $|l-s|$ for heavy rare earths. For example, Europium has $s=3$ and $l=3$ as effective spin and orbital momenta, respectively, and its ground state corresponds to $j_{0}=0$, which is clearly an entangled state, because it is a singlet. However, for Terbium, which has the same angular momenta, the $j=0$ state is the most energetic, and at room temperatures the entanglement contained in this state never contributes to the thermal entanglement. The main contribution to the witness for Terbium comes from the ground state $j_{0}=6$, which is highly degenerate, so non-entanglement of the mixture can be assumed. The same happens with the other light rare earths and their heavy counterparts.

Summarizing, the present study explores the thermal entanglement in rare earth ions. This quantum correlation was verified through an energy-based entanglement witness, constructed so as to be sensitive to entanglement not between ions, but between two variables of the same ion (spin and orbital angular momenta), through the spin-orbit interaction. Thus rare-earth ions present an internal degree of entanglement much more robust to temperature fluctuations than any other condensed matter systems composed of many particles.

It is interesting to note that in the light rare-earth ions entanglement survives up to very high temperatures, and the results show entanglement temperatures from 1758 K (for Ce) up to 3295 K (for Eu), while no entanglement could be found for heavy rare-earth ions at any temperature. The classification of these two rare-earth sub-groups depends on the filling of the $4f$ electron shell, and the consequences are the presence or absence of entanglement inside the ion. Thus we present a physical explanation for the origin of entanglement in these ions due to the distribution and occupation of the different energy levels that enhances our understanding of thermal entanglement in physical systems.

\acknowledgments We gratefully acknowledge CAPES, CNPq, FAPERJ and PROPPi-UFF for the financial support. We also gratefully acknowledge Mrs. K. Friedman for the careful reading of the manuscript. This work was performed as part of the Brazilian National Institute of Science and Technology for Quantum Information (INCT-IQ).

%

\end{document}